\documentstyle[12pt]{article}
\newcommand{ \R} {\mbox{\rm I$\!$R}}
\newcommand{ \C} {\mbox{\rm I$\!$C}}
\begin{document}

\title{Gauge and Einstein Gravity from\\
Non--Abelian Gauge Models on Noncommutative Spaces}
\author{Sergiu I.\ Vacaru \thanks{
e--mail: vacaru@lises.asm.md } \quad \\
%EndAName
\\
{\small Institute of Applied Physics, Academy of Sciences, }\\
{\small 5 Academy str., Chi\c sin\v au MD2028, Republic of Moldova }}
\date{September 22, 2000}
\maketitle

\begin{abstract}
Following the formalism of enveloping algebras and star product calculus we
formulate and analyze a model of gauge gravity on noncommutative spaces and
examine the conditions of its equivalence to general relativity. The
corresponding Seiber--Witten maps are established which allow the
definition of respective dynamics for a finite number of gravitational gauge
field components on noncommutative spaces.
\end{abstract}
\vskip0.5cm

%\tableofcontents
%%%%%%%%%%%%%%%%%%%%%%%%%%%%%%%%%%%%%%%%%%%%%%%%%%%%%%%%%%%%%%%%%%%%

\section{Introduction}

In the last years much work has been made in noncommutative extensions of
physical theories. It was not possible to formulate gauge theories on
noncommutative spaces \cite{cds,sw,js,mssw} with Lie algebra valued
infinitesimal transformations and with Lie algebra valued gauge fields. In
order to avoid the problem the authors of \cite{jssw} suggested to use
enveloping algebras of the Lie algebras for setting this type of gauge
theories and showed that in spite of the fact that such enveloping algebras
are infinite--dimensional one can restrict them in a way that it would be a
dependence on the Lie algebra valued parameters and the Lie algebra valued
gauge fields and their spacetime derivatives only.

A still presented drawback of noncommutative geometry and physics is that
there is not yet formulated a generally accepted approach to interactions of
elementary particles coupled to gravity. There are improved Connes--Lott and
Chamsedine--Connes models of nocommutative geometry \cite{cl} which yielded
action functionals typing together the gravitational and Yang--Mills
interactions and gauge bosons the Higgs sector (see also the approaches \cite
{hawkins} and, for an outline of recent results, \cite{majid}).

In this paper we follow the method of restricted enveloping algebras \cite
{js,jssw} and construct gauge gravitational theories by stating
corresponding structures with semisimple or nonsemisimple Lie algebras and
their extensions. We consider power series of generators for the affine and
non linear realized de Sitter gauge groups and compute the coefficient
functions of al the higher powers of the generators of the gauge group which
are functions of the coefficients of the first power. Such constructions are
based on the Seiberg--Witten map \cite{sw} and on the formalism of $*$%
--product formulation of the algebra \cite{w} when for functional objects,
being functions of commuting variables, there are associated some algebraic
noncommutative properties encoded in the $*$--product.

The concept of gauge theory on noncommutative spaces was introduced in a
geometric manner \cite{mssw} by defining the covariant coordinates without
speaking about derivatives and this formalism was developed for quantum
planes \cite{wz}. In this paper we shall prove the existence for
noncommutative spaces of gauge models of gravity which agrees with usual
gauge gravity theories being equivalent or extending the general relativity
theory (see works \cite{pd,ts} for locally isotropic spaces and
corresponding reformulations and generalizations respectively for
anholonomic frames \cite{vd} and locally anisotropic (super) spaces \cite{vg}%
) in the limit of commuting spaces.

\section{*--Products and Enveloping Algebras \newline in
Noncommutative Spaces}

For a noncommutative space the coordinates ${\hat u}^i,$ $(i=1,...,N)$
satisfy some noncommutative relations of type
\begin{equation}
\label{ncr}[{\hat u}^i,{\hat u}^j]=\left\{
\begin{array}{rcl}
& i\theta ^{ij}, & \theta ^{ij}\in
\C,\mbox{ canonical structure; } \\  & if_k^{ij}{\hat u}^k, & f_k^{ij}\in
\C,\mbox{ Lie structure; } \\  & iC_{kl}^{ij}{\hat u}^k{\hat u}^l, &
C_{kl}^{ij}\in \C ,\mbox{ quantum plane structure}
\end{array}
\right.
\end{equation}
where $\C$ denotes the complex number field.

The noncommutative space is modeled as the associative algebra of $\C;$\
this algebra is freely generated by the coordinates modulo ideal ${\cal R}$
generated by the relations (one accepts formal power series)\ ${\cal A}_u=%
\C[[{\hat u}^1,...,{\hat u}^N]]/{\cal R.}$ One restricts attention \cite
{jssw} to algebras having the (so--called, Poincare--Birkhoff--Witt)
property that any element of ${\cal A}_u$ is defined by its coefficient
function and vice versa,%
$$
\widehat{f}=\sum\limits_{L=0}^\infty f_{i_1,...,i_L}:{\hat u}^{i_1}\ldots {%
\hat u}^{i_L}:\quad \mbox{ when }\widehat{f}\sim \left\{ f_i\right\} ,
$$
where $:{\hat u}^{i_1}\ldots {\hat u}^{i_L}:$ denotes that the basis
elements satisfy some prescribed order (for instance, the normal order $%
i_1\leq i_2\leq \ldots \leq i_L,$ or, another example, are totally
symmetric). The algebraic properties are all encoded in the so--called
diamond $(\diamond )$ product which is defined by
$$
\widehat{f}\widehat{g}=\widehat{h}~\sim ~\left\{ f_i\right\} \diamond
\left\{ g_i\right\} =\left\{ h_i\right\} .
$$

In the mentioned approach to every function $f(u)=f(u^1,\ldots ,u^N)$ of
commuting variables $u^1,\ldots ,u^N$ one associates an element of algebra $%
\widehat{f}$ when the commuting variables are substituted by anticommuting
ones,
$$
f(u)=\sum f_{i_1\ldots i_L}u^1\cdots u^N\rightarrow \widehat{f}%
=\sum\limits_{L=0}^\infty f_{i_1,...,i_L}:{\hat u}^{i_1}\ldots {\hat u}%
^{i_L}:
$$
when the $\diamond $--product leads to a bilinear $*$--product of functions
(see details in \cite{mssw})%
$$
\left\{ f_i\right\} \diamond \left\{ g_i\right\} =\left\{ h_i\right\} \sim
\left( f*g\right) \left( u\right) =h\left( u\right) .
$$

The $*$--product is defined respectively for the cases (\ref{ncr})
$$
f*g=\left\{
\begin{array}{rcl}
\exp [{\frac i2}{\frac \partial {\partial u^i}}{\theta }^{ij}\frac \partial
{\partial {u^{\prime }}^j}]f(u)g(u^{\prime }){|}_{u^{\prime }\to u}, &
\mbox{  canonical structure;} &  \\
\exp [\frac i2u^kg_k(i\frac \partial {\partial u^{\prime }},i\frac \partial
{\partial u^{\prime \prime }})]f(u^{\prime })g(u^{\prime \prime }){|}%
_{u^{\prime \prime }\to u}^{u^{\prime }\to u}, & \mbox{  Lie structure;} &
\\
q^{{\frac 12}(-u^{\prime }{\frac \partial {\partial u^{\prime }}}v{\frac
\partial {\partial v}}+u{\frac \partial {\partial u}}v^{\prime }{\frac
\partial {\partial v^{\prime }}})}f(u,v)g(u^{\prime },v^{\prime }){|}%
_{v^{\prime }\to v}^{u^{\prime }\to u}, & \mbox{  quantum plane}, &
\end{array}
\right.
$$
where there are considered values of type%
\begin{eqnarray}
e^{ik_n\widehat{u}^n}e^{ip_{nl}\widehat{u}^n} &=&e^{i\{k_n+p_n+\frac
12g_n\left( k,p\right) \}\widehat{u}^n,}  \nonumber \\
g_n\left( k,p\right) & =& -k_ip_jf_{\ n}^{ij}+\frac 16k_ip_j\left(
p_k-k_k\right) f_{\ m}^{ij}f_{\ n}^{mk}+..., \label{gdecomp} \\
e^Ae^B &=&
e^{A+B+\frac 12[A,B]+\frac 1{12}\left( [A,[A,B]]+[B,[B,A]]\right)
}+... \nonumber
\end{eqnarray}
and for the coordinates on quantum (Manin) planes one holds the relation $%
uv=qvu.$

A non--abelian gauge theory on a noncommutative space is given by two
algebraic structures, the algebra ${\cal A}_u$ and a non--abelian Lie
algebra ${\cal A}_I$ of the gauge group with generators $I^1,...,I^S$ and
the relations
\begin{equation}
\label{commutators1}[I^{\underline{s}},I^{\underline{p}}]=if_{~\underline{t}%
}^{\underline{s}\underline{p}}I^{\underline{t}}.
\end{equation}
In this case both algebras are treated on the same footing and one denotes
the generating elements of the big algebra by $\widehat{u}^i,$%
\begin{eqnarray}
\widehat{z}^{\underline i} &=&
\{\widehat{u}^1,...,\widehat{u}^N,I^1,...,I^S\}, \nonumber \\
{\cal A}_z &=&\C[[\widehat{u}^1,...,\widehat{u}^{N+S}]]/{\cal R,}
\nonumber
\end{eqnarray}
and the $*$--product formalism is to be applied for the whole algebra ${\cal %
A}_z$ when there are considered functions of the commuting variables $u^i\
(i,j,k,...=1,...,N)$ and $I^{\underline{s}}\ (s,p,...=1,...,S).$

For instance, in the case of a canonical structure for the space variables $%
u^i$ we have
\begin{equation}
\label{csp1}(F*G)(u)=e^{\frac i2\left( \theta ^{ij}\frac \partial {\partial
u^{\prime i}}\frac \partial {\partial u^{\prime \prime j}}+t^sg_s\left(
i\frac \partial {\partial t^{\prime }},i\frac \partial {\partial t^{\prime
\prime }}\right) \right) }F\left( u^{\prime },t^{\prime }\right) G\left(
u^{\prime \prime },t^{\prime \prime }\right) \mid _{t^{\prime }\rightarrow
t,t^{\prime \prime }\rightarrow t}^{u^{\prime }\rightarrow u,u^{\prime
\prime }\rightarrow u}.
\end{equation}
This formalism was developed in \cite{jssw} for general Lie algebras. In
this paper we shall consider those cases when in the commuting limit one
obtains the gauge gravity and general relativity theories.

\section{Enveloping Algebras for \newline Gravitational Gauge
Connections}

To define gauge gravity theories on noncommutative space we first introduce
gauge fields as elements the algebra ${\cal A}_u$ that form representation
of the generator $I$--algebra for the de Sitter gauge group. For commutative
spaces it is known \cite{pd,ts,vg} that an equivalent reexpression of the
Einstein theory as a gauge like theory implies, for both locally isotropic
and anisotropic spacetimes, the nonsemisimplicity of the gauge group, which
leads to a nonvariational theory in the total space of the bundle of locally
adapted affine frames (to this class one belong the gauge Poincare theories;\ 
on metric--affine and gauge gravity models see original results and reviews
in \cite{ut}). By using auxililiary biliniear forms, instead  of degenerated
Killing form for the affine structural group,  on fiber spaces, the gauge
models of gravity can be formulated to be  variational. After projection on
the base spacetime, for the so--called  Cartan connection form, the
Yang--Mills equations transforms equivalently  into the Einstein equations
for general relativity \cite{pd}. A variational gauge gravitational theory
can be also formulated by using a minimal extension of the affine structural
group ${{\cal A}f}_{3+1}\left( {\R}\right) $ to the de Sitter gauge group $%
S_{10}=SO\left( 4+1\right) $ acting on ${\R}^{4+1}$ space.

\subsection{Nonlinear gauge theories of de Sitter group  \newline in
commutative spaces}

Let us consider the de Sitter space $\Sigma ^4$ as a hypersurface given by
the equations $\eta _{AB}u^Au^B=-l^2$ in the four dimensional flat space
enabled with diagonal metric $\eta _{AB},\eta _{AA}=\pm 1$ (in this section $%
A,B,C,...=1,2,...,5),$ where $\{u^A\}$ are global Cartesian coordinates in $%
\R^5;l>0$ is the curvature of de Sitter space. The de Sitter group $%
S_{\left( \eta \right) }=SO_{\left( \eta \right) }\left( 5\right) $ is
defined as the isometry group of $\Sigma ^5$--space with $6$ generators of
Lie algebra ${{\it s}o}_{\left( \eta \right) }\left( 5\right) $ satisfying
the commutation relations
\begin{equation}
\label{dsc}\left[ M_{AB},M_{CD}\right] =\eta _{AC}M_{BD}-\eta
_{BC}M_{AD}-\eta _{AD}M_{BC}+\eta _{BD}M_{AC}. %\eqno(3.34)%47
\end{equation}

Decomposing indices $A,B,...$ as $A=\left( \underline{\alpha },5\right)
,B=\left( \underline{\beta },5\right) ,...,$ the metric $\eta _{AB}$ as $%
\eta _{AB}=\left( \eta _{\underline{\alpha }\underline{\beta }},\eta
_{55}\right) ,$ and operators $M_{AB}$ as $M_{\underline{\alpha }\underline{%
\beta }}={\cal F}_{\underline{\alpha }\underline{\beta }}$ and $P_{%
\underline{\alpha }}=l^{-1}M_{5\underline{\alpha }},$ we can write (\ref{dsc}%
) as
\begin{eqnarray}
\left[ {\cal F}_{\underline{\alpha }\underline{\beta }},
{\cal F}_{\underline{\gamma }\underline{\delta }}\right] & = &
\eta _{\underline{\alpha }\underline{\gamma }}
{\cal F}_{\underline{\beta }\underline{\delta }}-
\eta _{\underline{\beta }\underline{\gamma }}
{\cal F}_{\underline{\alpha }\underline{\delta }}+
\eta _{\underline{\beta }\underline{\delta }}
{\cal F}_{\underline{\alpha }\underline{\gamma }}-
\eta _{\underline{\alpha }\underline{\delta }}
{\cal F}_{\underline{\beta }\underline{\gamma }}, \label{dsca} \\
\left[ P_{\underline{\alpha }},P_{\underline{\beta }}\right] & = &
-l^{-2}{\cal F}%
_{\underline{\alpha }\underline{\beta }},\quad
\left[ P_{\underline{\alpha }},
{\cal F}_{\underline{\beta }\underline{\gamma }}\right] = %
\eta _{\underline{\alpha }\underline{\beta }}P_{\underline{\gamma }}-
\eta _{\underline{\alpha }\underline{\gamma }}P_{\underline{\beta }},
\nonumber
\end{eqnarray}
where we have indicated the possibility to decompose the Lie algebra ${{\it s%
}o}_{\left( \eta \right) }\left( 5\right) $ into a direct sum, ${{\it s}o}%
_{\left( \eta \right) }\left( 5\right) ={{\it s}o}_{\left( \eta \right)
}(4)\oplus V_4,$ where $V_4$ is the vector space stretched on vectors $P_{%
\underline{\alpha }}.$ We remark that $\Sigma ^4=S_{\left( \eta \right)
}/L_{\left( \eta \right) },$ where $L_{\left( \eta \right) }=SO_{\left( \eta
\right) }\left( 4\right) .$ For $\eta _{AB}=diag\left( 1,-1,-1,-1\right) $
and $S_{10}=SO\left( 1,4\right) ,L_6=SO\left( 1,3\right) $ is the group of
Lorentz rotations.

In this paper the generators $I^{\underline{a}}$ and structure constants $%
f_{~\underline{t}}^{\underline{s}\underline{p}}$ from (\ref{commutators1})
are parametrized just to obtain de Sitter generators and commutations  (\ref
{dsca}).

The action of the group $S_{\left( \eta \right) }$ can be realized by using $%
4\times 4$ matrices with a parametrization distinguishing subgroup $%
L_{\left( \eta \right) }:$%
\begin{equation}
\label{parametriz}B=bB_L, %\eqno(3.35)
\end{equation}
where%
$$
B_L=\left(
\begin{array}{cc}
L & 0 \\
0 & 1
\end{array}
\right) ,
$$
$L\in L_{\left( \eta \right) }$ is the de Sitter bust matrix transforming
the vector $\left( 0,0,...,\rho \right) \in {\R}^5$ into the arbitrary point
$\left( V^1,V^2,...,V^5\right) \in \Sigma _\rho ^5\subset {\cal R}^5$ with
curvature $\rho,$ $(V_A V^A=-\rho ^2, V^A=t^A\rho ).$ Matrix $b$ can be
expressed as
$$
b=\left(
\begin{array}{cc}
\delta _{\quad \underline{\beta }}^{\underline{\alpha }}+\frac{t^{\underline{%
\alpha }}t_{\underline{\beta }}}{\left( 1+t^5\right) } & t^{
\underline{\alpha }} \\ t_{\underline{\beta }} & t^5
\end{array}
\right) .
$$

The de Sitter gauge field is associated with a ${{\it s}o}_{\left( \eta
\right) }\left( 5\right) $--valued connection 1--form
\begin{equation}
\label{dspot}\widetilde{\Omega }=\left(
\begin{array}{cc}
\omega _{\quad \underline{\beta }}^{\underline{\alpha }} & \widetilde{\theta
}^{\underline{\alpha }} \\ \widetilde{\theta }_{\underline{\beta }} & 0
\end{array}
\right) ,% \eqno(3.36)%49
\end{equation}
where $\omega _{\quad \underline{\beta }}^{\underline{\alpha }}\in
so(4)_{\left( \eta \right) },$ $\widetilde{\theta }^{\underline{\alpha }}\in
{\cal R}^4,\widetilde{\theta }_{\underline{\beta }}\in \eta _{\underline{%
\beta }\underline{\alpha }}\widetilde{\theta }^{\underline{\alpha }}.$

Because $S_{\left( \eta \right) }$--transforms mix $\omega _{\quad
\underline{\beta }}^{\underline{\alpha }}$ and $\widetilde{\theta }^{%
\underline{\alpha }}$ fields in (\ref{dspot}) %(3.36)
(the introduced para\-met\-ri\-za\-ti\-on is invariant on action on $%
SO_{\left( \eta \right) }\left( 4\right) $ group we cannot identify $\omega
_{\quad \underline{\beta }}^{\underline{\alpha }}$ and $\widetilde{\theta }^{%
\underline{\alpha }},$ respectively, with the connection $\Gamma _{~\beta
\gamma }^\alpha $ and the fundamental form $\chi ^\alpha $ in a
metric--affine spacetime. To avoid this difficulty we consider \cite{ts} a
nonlinear gauge realization of the de Sitter group $S_{\left( \eta \right)
}, $ namely, we introduce into consideration the nonlinear gauge field
\begin{equation}
\label{npot}\Gamma =b^{-1}{\widetilde{\Omega }}b+b^{-1}db=\left(
\begin{array}{cc}
\Gamma _{~\underline{\beta }}^{\underline{\alpha }} & \theta ^{
\underline{\alpha }} \\ \theta _{\underline{\beta }} & 0
\end{array}
\right) ,%\eqno(3.37)%50
\end{equation}
where
\begin{eqnarray}
\Gamma _{\quad \underline{\beta }}^{\underline{\alpha }} &=&\omega _{\quad
\underline{\beta }}^{\underline{\alpha }}-
\left( t^{\underline{\alpha }}Dt_{\underline{\beta }} -
t_{\underline{\beta }}Dt^{\underline{\alpha }}\right)
/\left( 1+t^5\right) , \nonumber \\
\theta ^{\underline{\alpha }} & = &
t^5\widetilde{\theta }^{\underline{\alpha }}+Dt^{\underline{\alpha }}-
 t^{\underline{\alpha }}\left( dt^5+
 \widetilde{\theta }_{\underline{\gamma }}t^{\underline{\gamma }}\right) /
\left( 1+t^5\right) ,  \nonumber \\
Dt^{\underline{\alpha }} &=& dt^{\underline{\alpha }}+
\omega _{\quad \underline{\beta }}^{\underline{\alpha }}
t^{\underline{\beta }}. \nonumber
\end{eqnarray}

The action of the group $S\left( \eta \right) $ is nonlinear, yielding
transforms
$$
\Gamma ^{\prime }=L^{\prime }\Gamma \left( L^{\prime }\right)
^{-1}+L^{\prime }d\left( L^{\prime }\right) ^{-1},\theta ^{\prime }=L\theta
,
$$
where the nonlinear matrix--valued function $L^{\prime }=L^{\prime }\left(
t^\alpha ,b,B_T\right) $ is defined from $B_b=b^{\prime }B_{L^{\prime }}$
(see the parametrization (\ref{parametriz})). The de Sitter algebra with
generators (\ref{dsca}) and nonlinear gauge transforms of type (\ref{npot})
is denoted ${\cal A}_I^{(dS)}.$

\subsection{Enveloping nonlinear de Sitter \newline %
algebra valued connection}

Let now consider a noncommutative space. In this case the gauge fields are
elements of the algebra $\widehat{\psi }\in {\cal A}_I^{(dS)}$ that form the
nonlinear representation of the de Sitter algebra ${{\it s}o}_{\left( \eta
\right) }\left( 5\right) $ when the whole algebra is denoted ${\cal A}%
_z^{(dS)}$ Under a nonlinear de Sitter transformation the elements transform
as follows%
$$
\delta \widehat{\psi }=i\widehat{\gamma }\widehat{\psi },\widehat{\psi }\in
{\cal A}_u,\widehat{\gamma }\in {\cal A}_z^{(dS)},
$$
So, the action of the generators (\ref{dsca}) on $\widehat{\psi }$ is
defined as this element is supposed to form a nonlinear representation of $%
{\cal A}_I^{(dS)}$ and, in consequence, $\delta \widehat{\psi }\in {\cal A}%
_u $ despite $\widehat{\gamma }\in {\cal A}_z^{(dS)}.$ It should be
emphasized that independent of a representation the object $\widehat{\gamma }
$ takes values in enveloping de Sitter algebra and not in a Lie algebra as
would be for commuting spaces. The same holds for the connections that we
introduce (similarly to \cite{mssw}) in order to define covariant coordinates%
$$
\widehat{U}^\nu =\widehat{u}^v+\widehat{\Gamma }^\nu ,\widehat{\Gamma }^\nu
\in {\cal A}_z^{(dS)}.
$$

The values $\widehat{U}^\nu \widehat{\psi }$ transforms covariantly, $\delta
\widehat{U}^\nu \widehat{\psi }=i\widehat{\gamma }\widehat{U}^\nu \widehat{%
\psi },$ if and only if the connection $\widehat{\Gamma }^\nu $ satisfies
the transformation law of the enveloping nonlinear realized de Sitter
algerba,%
$$
\delta \widehat{\Gamma }^\nu \widehat{\psi }=-i[\widehat{u}^v,\widehat{%
\gamma }]+i[\widehat{\gamma },\widehat{\Gamma }^\nu ],
$$
where $\delta \widehat{\Gamma }^\nu \in {\cal A}_z^{(dS)}.$ The enveloping
algebra--valued connection has infinitely many component fields.
Nevertheless, it was shown that all the component fields can be induced from
a Lie algebra--valued connection by a Seiberg--Witten map (\cite{sw,js,jssw}
and \cite{bsst} for $SO(n)$ and $Sp(n)).$ In this subsection we show that
similar constructions could be proposed for nonlinear realizations of de
Sitter algebra when the transformation of the connection is considered%
$$
\delta \widehat{\Gamma }^\nu =-i[u^\nu ,^{*}~\widehat{\gamma }]+i[\widehat{%
\gamma },^{*}~\widehat{\Gamma }^\nu ].
$$
For simplicity, we treat in more detail the canonical case with the star
product (\ref{csp1}). The first term in the variation $\delta \widehat{%
\Gamma }^\nu $ gives
$$
-i[u^\nu ,^{*}~\widehat{\gamma }]=\theta ^{\nu \mu }\frac \partial {\partial
u^\mu }\gamma .
$$
Assuming that the variation of $\widehat{\Gamma }^\nu =\theta ^{\nu \mu
}Q_\mu $ starts with a linear term in $\theta $ we have%
$$
\delta \widehat{\Gamma }^\nu =\theta ^{\nu \mu }\delta Q_\mu ,~\delta Q_\mu
=\frac \partial {\partial u^\mu }\gamma +i[\widehat{\gamma },^{*}~Q_\mu ].
$$
We follow the method of calculation from the papers \cite{mssw,jssw} and
expand the star product (\ref{csp1}) in $\theta $ but not in $g_a$ and find
to first order in $\theta ,$%
\begin{eqnarray}
\gamma &=&
\gamma _{\underline{a}}^1I^{\underline{a}}+\gamma _{\underline{a} %
\underline{b}}^1I^{\underline{a}}I^{\underline{b}}+..., \label{series}\\%
Q_\mu &=& q_{\mu ,\underline{a}}^1I^{\underline{a}}+q_{\mu ,\underline{a} %
\underline{b}}^2I^{\underline{a}}I^{\underline{b}}+... \nonumber %
\end{eqnarray}
where $\gamma _{\underline{a}}^1$ and $q_{\mu ,\underline{a}}^1$ are of
order zero in $\theta $ and $\gamma _{\underline{a}\underline{b}}^1$ and $%
q_{\mu ,\underline{a}\underline{b}}^2$ are of second order in $\theta .$ The
expansion in $I^{\underline{b}}$ leads to an expansion in $g_a$ of the $*$%
--product because the higher order $I^{\underline{b}}$--derivatives vanish.
For de Sitter case as $I^{\underline{b}}$ we take the generators (\ref{dsca}%
), see commutators (\ref{commutators1}), with the corresponding de Sitter
structure constants $f_{~\underline{d}}^{\underline{b}\underline{c}}\simeq
f_{~\underline{\beta }}^{\underline{\alpha }\underline{\beta }}$ (in our
further identifications with spacetime objects like frames and connections
we shall use Greeck indices).

The result of calculation of variations of (\ref{series}), by using $g_a$ to
the order given in (\ref{gdecomp}), is%
\begin{eqnarray}
\delta q_{\mu ,\underline{a}}^1
 &=&\frac{\partial \gamma _{\underline{a}}^1}{\partial u^\mu }- %
f_{~\underline{a}}^{\underline{b}\underline{c}} %
\gamma _{\underline{b}}^1q_{\mu ,\underline{c}}^1, \nonumber \\  %
\delta Q_\tau &=& \theta ^{\mu \nu } %
\partial _\mu \gamma _{\underline{a}}^1\partial _\nu q_{\tau , %
\underline{b}}^1I^{\underline{a}}I^{\underline{b}}+...,  \nonumber \\ %
\delta q_{\mu ,\underline{a}\underline{b}}^2 &=&  %
\partial _\mu \gamma _{\underline{a}\underline{b}}^2 %
 -\theta ^{\nu \tau }\partial _\nu \gamma _{\underline{a}}^1 %
\partial _\tau q_{\mu ,\underline{b}}^1- %
2f_{~\underline{a}}^{\underline{b}\underline{c}} %
\{\gamma _{\underline{b}}^1q_{\mu ,\underline{c}\underline{d}}^2+ %
\gamma _{\underline{b}\underline{d}}^2q_{\mu ,\underline{c}}^1\}. %
 \nonumber
\end{eqnarray}

Next we introduce the objects $\varepsilon ,$ taking the values in de Sitter
Lie algebra and $W_\mu ,$ being enveloping de Sitter algebra valued,%
$$
\varepsilon =\gamma _{\underline{a}}^1I^{\underline{a}}\mbox{ and }W_\mu
=q_{\mu ,\underline{a}\underline{b}}^2I^{\underline{a}}I^{\underline{b}}
$$
with the variation $\delta W_\mu $ satisfying the equation \cite{mssw,jssw}
\begin{equation}
\label{vareq}\delta W_\mu =\partial _\mu (\gamma _{\underline{a}\underline{b}%
}^2I^{\underline{a}}I^{\underline{b}})-\frac 12\theta ^{\tau \lambda
}\{\partial _\tau \varepsilon ,\partial _\lambda q_\mu \}+i[\varepsilon
,W_\mu ]+i[(\gamma _{\underline{a}\underline{b}}^2I^{\underline{a}}I^{%
\underline{b}}),q_\nu ].
\end{equation}
The equation (\ref{vareq}) has the solution (found in (\cite{mssw,sw}))%
\begin{eqnarray}
\gamma _{\underline{a}\underline{b}}^2 &=&
\frac 12\theta ^{\nu \mu }(\partial
_\nu \gamma _{\underline{a}}^1)q_{\mu ,\underline{b}}^1, \nonumber \\
q_{\mu ,\underline{a}\underline{b}}^2 &=&
-\frac 12\theta ^{\nu \tau }q_{\nu ,\underline{a}}^1
\left( \partial _\tau q_{\mu ,\underline{b}}^1+ %
R_{\tau \mu ,\underline{b}}^1\right) \nonumber %
\end{eqnarray}
where
$$
R_{\tau \mu ,\underline{b}}^1=\partial _\tau q_{\mu ,\underline{b}%
}^1-\partial _\mu q_{\tau ,\underline{b}}^1+f_{~\underline{d}}^{\underline{e}%
\underline{c}}q_{\tau ,\underline{e}}^1q_{\mu ,\underline{e}}^1
$$
could be identified with the coefficients ${\cal R}_{\quad \underline{\beta }%
\mu \nu }^{\underline{\alpha }}$ of de Sitter nonlinear gauge gravity
curvature (see formula (2a) from the Appendix) if in the commutative limit $%
q_{\mu ,\underline{b}}^1\simeq \left(
\begin{array}{cc}
\Gamma _{\quad \underline{\beta }}^{\underline{\alpha }} & l_0^{-1}\chi ^{
\underline{\alpha }} \\ l_0^{-1}\chi _{\underline{\beta }} & 0
\end{array}
\right) $ (see (1a)).

The below presented procedure can be generalized to all the higher powers of
$\theta $ \cite{jssw}.

\section{Noncommutative Gauge Gravity Covariant Dynamics}

\subsection{First order corrections to gravitational curvature}

The constructions from the previous section are summarized by the conclusion
that the de Sitter algebra valued object $\varepsilon =\gamma _{\underline{a}%
}^1\left( u\right) I^{\underline{a}}$ determines all the terms in the
enveloping algebra%
$$
\gamma =\gamma _{\underline{a}}^1I^{\underline{a}}+\frac 14\theta ^{\nu \mu
}\partial _\nu \gamma _{\underline{a}}^1\ q_{\mu ,\underline{b}}^1\left( I^{%
\underline{a}}I^{\underline{b}}+I^{\underline{b}}I^{\underline{a}}\right)
+...
$$
and the gauge transformations are defined by $\gamma _{\underline{a}%
}^1\left( u\right) $ and $q_{\mu ,\underline{b}}^1(u),$ when
$$
\delta _{\gamma ^1}\psi =i\gamma \left( \gamma ^1,q_\mu ^1\right) *\psi .
$$
For de Sitter enveloping algebras one holds the general formula for
compositions of two transformations%
$$
\delta _\gamma \delta _\varsigma -\delta _\varsigma \delta _\gamma =\delta
_{i(\varsigma *\gamma -\gamma *\varsigma )}
$$
which holds also for the restricted transformations defined by $\gamma ^1,$%
$$
\delta _{\gamma ^1}\delta _{\varsigma ^1}-\delta _{\varsigma ^1}\delta
_{\gamma ^1}=\delta _{i(\varsigma ^1*\gamma ^1-\gamma ^1*\varsigma ^1)}.
$$

Applying the formula (\ref{csp1}) we computer%
\begin{eqnarray}
[\gamma ,^{*}\zeta ]
 &=&i\gamma _{\underline{a}}^1\zeta _{\underline{b}}^1 %
f_{~\underline{c}}^{\underline{a}\underline{b}}I^{\underline{c}}+%
\frac i2\theta ^{\nu \mu }\{\partial _v\left( \gamma _{\underline{a}}^1%
\zeta _{\underline{b}}^1 %
f_{~\underline{c}}^{\underline{a}\underline{b}}\right) %
 q_{\mu ,\underline{c}} \nonumber \\  &{}&
+\left( \gamma _{\underline{a}}^1\partial _v\zeta _{\underline{b}}^1- %
\zeta _{\underline{a}}^1\partial _v\gamma _{\underline{b}}^1\right) %
 q_{\mu ,\underline{b}}f_{~\underline{c}}^{\underline{a}\underline{b}}+%
 2\partial _v %
\gamma _{\underline{a}}^1\partial _\mu \zeta _{\underline{b}}^1\} %
I^{\underline{d}}I^{\underline{c}}. \nonumber %
\end{eqnarray}
Such commutators could be used for definition of tensors \cite{mssw}
\begin{equation}
\label{tensor1}\widehat{S}^{\mu \nu }=[\widehat{U}^\mu ,\widehat{U}^\nu ]-i%
\widehat{\theta }^{\mu \nu },
\end{equation}
where $\widehat{\theta }^{\mu \nu }$ is respectively stated for the
canonical, Lie and quantum plane structures. Under the general enveloping
algebra one holds the transform%
$$
\delta \widehat{S}^{\mu \nu }=i[\widehat{\gamma },\widehat{S}^{\mu \nu }].
$$
For instance, the canonical case is characterized by%
\begin{eqnarray}
S^{\mu \nu } &=&
i\theta ^{\mu \tau }\partial _\tau \Gamma ^\nu -i\theta ^{\nu
\tau }\partial _\tau \Gamma ^\mu +\Gamma ^\mu *\Gamma ^\nu -\Gamma ^\nu
*\Gamma ^\mu  \nonumber \\
& = &
\theta ^{\mu \tau }\theta ^{\nu \lambda }\{\partial _\tau Q_\lambda
-\partial _\lambda Q_\tau +Q_\tau *Q_\lambda -Q_\lambda *Q_\tau \}.
\nonumber
\end{eqnarray}
By introducing the gravitational gauge strength (curvature)
\begin{equation}
\label{qcurv}R_{\tau \lambda }=\partial _\tau Q_\lambda -\partial _\lambda
Q_\tau +Q_\tau *Q_\lambda -Q_\lambda *Q_\tau ,
\end{equation}
which could be treated as a noncommutative extension of de Sitter nonlinear
gauge gravitational curvature (2a), one computers
$$
R_{\tau \lambda ,\underline{a}}=R_{\tau \lambda ,\underline{a}}^1+\theta
^{\mu \nu }\{R_{\tau \mu ,\underline{a}}^1R_{\lambda \nu ,\underline{b}%
}^1-\frac 12q_{\mu ,\underline{a}}^1\left[ (D_\nu R_{\tau \lambda ,%
\underline{b}}^1)+\partial _\nu R_{\tau \lambda ,\underline{b}}^1\right]
\}I^{\underline{b}},
$$
where the gauge gravitation covariant derivative is introduced,%
$$
(D_\nu R_{\tau \lambda ,\underline{b}}^1)=\partial _\nu R_{\tau \lambda ,%
\underline{b}}^1+q_{\nu ,\underline{c}}R_{\tau \lambda ,\underline{d}}^1f_{~%
\underline{b}}^{\underline{c}\underline{d}}.
$$
Following the gauge transformation laws for $\gamma $ and $q^1$ we find
$$
\delta _{\gamma ^1}R_{\tau \lambda }^1=i\left[ \gamma ,^{*}R_{\tau \lambda
}^1\right]
$$
with the restricted form of $\gamma .$

Such formulas were proved in references \cite{jssw,sw} for usual gauge
(nongravitational) fields. Here we reconsidered them for gravitational gauge
fields.

\subsection{Gauge covariant gravitational dynamics}

Following the nonlinear realization of de Sitter algebra and the $*$%
--formalism we can formulate a dynamics of noncommutative spaces.
Derivatives can be introduced in such a way that one does not obtain new
relations for the coordinates. In this case a Leibniz rule can be defined
\cite{jssw} that
$$
\widehat{\partial }_\mu \widehat{u}^\nu =\delta _\mu ^\nu +d_{\mu \sigma
}^{\nu \tau }\ \widehat{u}^\sigma \ \widehat{\partial }_\tau
$$
where the coefficients $d_{\mu \sigma }^{\nu \tau }=\delta _\sigma ^\nu
\delta _\mu ^\tau $ are chosen to have not new relations when $\widehat{%
\partial }_\mu $ acts again to the right hand side. In consequence one holds
the $*$--derivative formulas
$$
\partial _\tau *f=\frac \partial {\partial u^\tau }f+f*\partial _\tau ,
$$
$$
[\partial _l,^{*}(f*g)]=\left( [\partial _l,^{*}f]\right) *g+f*\left(
[\partial _l,^{*}g]\right)
$$
and the Stokes theorem%
$$
\int [\partial _l,f]=\int d^Nu[\partial _l,^{*}f]=\int d^Nu\frac \partial
{\partial u^l}f=0,
$$
where, for the canonical structure, the integral is defined,%
$$
\int \widehat{f}=\int d^Nuf\left( u^1,...,u^N\right) .
$$

An action can be introduced by using such integrals. For instance, for a
tensor of type (\ref{tensor1}), when%
$$
\delta \widehat{L}=i\left[ \widehat{\gamma },\widehat{L}\right] ,
$$
we can define a gauge invariant action%
$$
W=\int d^Nu\ Tr\widehat{L},~\delta W=0,
$$
were the trace has to be taken for the group generator.

For the nonlinear de Sitter gauge gravity a proper action is
$$
L=\frac 14R_{\tau \lambda }R^{\tau \lambda },
$$
where $R_{\tau \lambda }$ is defined by (\ref{qcurv}) (in the commutative
limit we shall obtain the connection (1a)). In this case the dynamic of
noncommutative space is entirely formulated in the framework of quantum
field theory of gauge fields. The method works for matter fields as well to
restrictions to the general relativity theory (see references \cite{ts,pd}
 and the Appendix).

\section*{Appendix:\ De Sitter Nonlinear Gauge Gravity
 and General Relativity}

Let us consider the de Sitter nonlinear gauge gravitational connection (\ref
{npot}) rewritten in the form
$$
\Gamma =\left(
\begin{array}{cc}
\Gamma _{\quad \underline{\beta }}^{\underline{\alpha }} & l_0^{-1}\chi ^{
\underline{\alpha }} \\ l_0^{-1}\chi _{\underline{\beta }} & 0
\end{array}
\right) \eqno(1a)
$$
where
\begin{eqnarray}
\Gamma _{\quad \underline{\beta }}^{\underline{\alpha }} &=&\Gamma _{\quad
\underline{\beta }\mu }^{\underline{\alpha }}\delta u^\mu , \nonumber \\
\Gamma _{\quad \underline{\beta }\mu }^{\underline{\alpha }} &=&
\chi _{\quad \alpha }^{\underline{\alpha }}
\chi _{\quad \beta }^{\underline{\beta }}\Gamma _{\quad \beta \gamma
}^\alpha +\chi _{\quad \alpha }^{\underline{\alpha }}\delta _\mu \chi _{\quad
\underline{\beta }}^\alpha , \nonumber \\
\chi ^{\underline{\alpha }} &=& \chi _{\quad \mu }^{\underline{\alpha }}
\delta u^\mu , \nonumber
\end{eqnarray}
and
$$
G_{\alpha \beta }=\chi _{\quad \alpha }^{\underline{\alpha }}\chi _{\quad
\beta }^{\underline{\beta }}\eta _{\underline{\alpha }\underline{\beta }},
$$
 $\eta _{\underline{\alpha }\underline{\beta }}=\left(
1,-1,...,-1\right) $ and $l_0$ is a dimensional constant.

The curvature of (1a),
$$
{\cal R}^{(\Gamma )}=d\Gamma +\Gamma \bigwedge \Gamma ,
$$
can be written
$$
{\cal R}^{(\Gamma )}=\left(
\begin{array}{cc}
{\cal R}_{\quad \underline{\beta }}^{\underline{\alpha }}+l_0^{-1}\pi _{%
\underline{\beta }}^{\underline{\alpha }} & l_0^{-1}T^{
\underline{\alpha }} \\ l_0^{-1}T^{\underline{\beta }} & 0
\end{array}
\right) ,\eqno(2a)
$$
where
$$
\pi _{\underline{\beta }}^{\underline{\alpha }}=\chi ^{\underline{\alpha }%
}\bigwedge \chi _{\underline{\beta }},{\cal R}_{\quad \underline{\beta }}^{%
\underline{\alpha }}=\frac 12{\cal R}_{\quad \underline{\beta }\mu \nu }^{%
\underline{\alpha }}\delta u^\mu \bigwedge \delta u^\nu ,
$$
and
$$
{\cal R}_{\quad \underline{\beta }\mu \nu }^{\underline{\alpha }}=\chi _{%
\underline{\beta }}^{\quad \beta }\chi _\alpha ^{\quad \underline{\alpha }%
}R_{\quad \beta _{\mu \nu }}^\alpha .
$$
with $R_{\quad \beta {\mu \nu }}^\alpha $ being the metric--affine (for
Einstein--Cartan--Weyl spaces), or (pseudo) Riemannian curvature. The de
Sitter gauge group is semisimple and we are able to construct a variational
gauge gravitational theory with the Lagrangian
$$
L=L_{\left( G\right) }+L_{\left( m\right) }
$$
where the gauge gravitational Lagrangian is defined
$$
L_{\left( G\right) }=\frac 1{4\pi }Tr\left( {\cal R}^{(\Gamma )}\bigwedge *_G%
{\cal R}^{(\Gamma )}\right) ={\cal L}_{\left( G\right) }\left| G\right|
^{1/2}\delta ^4u,
$$
with
$$
{\cal L}_{\left( G\right) }=\frac 1{2l^2}T_{\quad \mu \nu }^{\underline{%
\alpha }}T_{\underline{\alpha }}^{\quad \mu \nu }+\frac 1{8\lambda }{\cal R}%
_{\quad \underline{\beta }\mu \nu }^{\underline{\alpha }}{\cal R}_{\quad
\underline{\alpha }}^{\underline{\beta }\quad \mu \nu }-\frac 1{l^2}\left( {%
\overleftarrow{R}}\left( \Gamma \right) -2\lambda _1\right) ,\eqno(3a)
$$
$\delta ^4u$ being the volume element, $T_{\quad \mu \nu }^{\underline{%
\alpha }}=\chi _{\quad \alpha }^{\underline{\alpha }}T_{\quad \mu \nu
}^\alpha $ (the gravitational constant $l^2$ in (3a) satisfies the relations
$l^2=2l_0^2\lambda ,\lambda _1=-3/l_0),\quad Tr$ denotes the trace on $%
\underline{\alpha },\underline{\beta }$ indices, and the matter field
Lagrangian is defined
$$
L_{\left( m\right) }=-1\frac 12Tr\left( \Gamma \bigwedge *_G{\cal I}\right) =%
{\cal L}_{\left( m\right) }\left| G\right| ^{1/2}\delta ^nu,
$$
where
$$
{\cal L}_{\left( m\right) }=\frac 12\Gamma _{\quad \underline{\beta }\mu }^{%
\underline{\alpha }}S_{\quad \alpha }^{\underline{\beta }\quad \mu
}-t_{\quad \underline{\alpha }}^\mu l_{\quad \mu }^{\underline{\alpha }}.%
\eqno(4a)
$$
The matter field source ${\cal J}$ is obtained as a variational derivation
of ${\cal L}_{\left( m\right) }$ on $\Gamma $ and is parametrized as
$$
{\cal J}=\left(
\begin{array}{cc}
S_{\quad \underline{\beta }}^{\underline{\alpha }} & -l_0t^{
\underline{\alpha }} \\ -l_0t_{\underline{\beta }} & 0
\end{array}
\right) \eqno(5a)
$$
with $t^{\underline{\alpha }}=t_{\quad \mu }^{\underline{\alpha }}\delta
u^\mu $ and $S_{\quad \underline{\beta }}^{\underline{\alpha }}=S_{\quad
\underline{\beta }\mu }^{\underline{\alpha }}\delta u^\mu $ being
respectively the canonical tensors of energy--momentum and spin density.

Varying the action
$$
S=\int \delta ^4u\left( {\cal L}_{\left( G\right) }+{\cal L}_{\left(
m\right) }\right)
$$
on the $\Gamma $--variables (1a), we obtain the gauge--gravitational field
equations:%
$$
d\left( *{\cal R}^{(\Gamma )}\right) +\Gamma \bigwedge \left( *{\cal R}%
^{(\Gamma )}\right) -\left( *{\cal R}^{(\Gamma )}\right) \bigwedge \Gamma
=-\lambda \left( *{\cal J}\right) ,\eqno(6a)
$$
were the Hodge operator $*$ is used.

Specifying the variations on $\Gamma _{\quad \underline{\beta }}^{\underline{%
\alpha }}$ and $\chi $--variables, we rewrite (6a)
\begin{eqnarray}
\widehat{{\cal D}}\left( *{\cal R}^{(\Gamma )}\right) +\frac{2\lambda }{l^2}%
\left( \widehat{{\cal D}}\left( *\pi \right) +\chi \bigwedge \left(
*T^T\right) -\left( *T\right) \bigwedge \chi ^T\right) & = & -\lambda \left(
*S\right) , \nonumber \\
\widehat{{\cal D}}\left( *T\right) -\left( *{\cal R}^{(\Gamma )}\right)
\bigwedge \chi -\frac{2\lambda }{l^2}\left( *\pi \right) \bigwedge \chi
 & = & \frac{l^2}2\left( *t+\frac 1\lambda *\tau \right) , \nonumber
\end{eqnarray}
where
\begin{eqnarray}
T^t &=&
\{T_{\underline{\alpha }}=\eta _{\underline{\alpha }     %
\underline{\beta }}T^{\underline{\beta }},~T^{\underline{\beta }}=   %
\frac 12T_{\quad \mu \nu }^{\underline{\beta }}\delta u^\mu \bigwedge %
\delta u^\nu \}, \nonumber \\ %
\chi ^T &=&   %
\{\chi _{\underline{\alpha }}=\eta _{\underline{\alpha }\underline{\beta }}%
\chi ^{\underline{\beta }},~\chi ^{\underline{\beta }}= %
\chi _{\quad \mu }^{\widehat{\beta }}\delta u^\mu \}, %
\qquad \widehat{{\cal D}}=d+\widehat{\Gamma },  \nonumber
\end{eqnarray}
($\widehat{\Gamma }$ acts as $\Gamma _{\quad \underline{\beta }\mu }^{%
\underline{\alpha }}$ on indices $\underline{\gamma },\underline{\delta }%
,... $ and as $\Gamma _{\quad \beta \mu }^\alpha $ on indices $\gamma
,\delta ,...).$ The value $\tau $ defines the energy--momentum tensor of the
gauge gravitational field $\widehat{\Gamma }:$%
$$
\tau _{\mu \nu }\left( \widehat{\Gamma }\right) =\frac 12Tr\left( {\cal R}%
_{\mu \alpha }{\cal R}_{\quad \nu }^\alpha -\frac 14{\cal R}_{\alpha \beta }%
{\cal R}^{\alpha \beta }G_{\mu \nu }\right) .
$$

Equations (6a) (or equivalently (7a)) make up the complete system of
variational field equations for nonlinear de Sitter gauge gravity.

We note that we can obtain a nonvariational Poincare gauge gravitational
theory if we consider the contraction of the gauge potential (1a) to a
potential with values in the Poincare Lie algebra
$$
\Gamma =\left(
\begin{array}{cc}
\Gamma _{\quad \widehat{\beta }}^{\widehat{\alpha }} & l_0^{-1}\chi ^{
\widehat{\alpha }} \\ l_0^{-1}\chi _{\widehat{\beta }} & 0
\end{array}
\right) \rightarrow \Gamma =\left(
\begin{array}{cc}
\Gamma _{\quad \widehat{\beta }}^{\widehat{\alpha }} & l_0^{-1}\chi ^{
\widehat{\alpha }} \\ 0 & 0
\end{array}
\right) .\eqno(7a)
$$
A similar gauge potential was considered in the formalism of linear and
affine frame bundles on curved spacetimes by Popov and Dikhin \cite{pd}.
They treated (7a) as the Cartan connection form for affine gauge like
gravity and by using 'pure' geometric methods proved that the Yang--Mills
equations of their theory are equivalent, after projection on the base, to
the Einstein equations. The main conclusion for a such approach to Einstein
gravity is that this theory admits an equivalent formulation as gauge model
but with nonsemisimple structural gauge groups. In order to have a
variational theory on the total bundle space it is necessary to introduce an
auxiliary bilinear form on the typical fiber, instead of degenerated Killing
form; the coefficients of auxiliary form disappear after pojection on the
base. An alternative variant is to consider a gauge gravitational theory
when the gauge  group was minimally extended to the de\ Sitter one with
nondegenerated Killing form. The nonlinear realizations have to be
introduced if we wont to consider in a common fashion both the frame
(tetradic) and connection components included as the coefficients of the
potential (1a).

%\newpage

\end{document}